\documentclass[lettersize,journal]{IEEEtran}

\usepackage{amsmath,amsfonts}
\DeclareMathOperator*{\argmax}{argmax}
\usepackage{algorithmic}
\usepackage{algorithm}
\usepackage{array}
\usepackage{cite}
\usepackage{hyperref}
\usepackage[capitalize]{cleveref}
\usepackage{graphicx}
\usepackage{mathtools}
\usepackage[caption=false,font=normalsize,labelfont=sf,textfont=sf]{subfig}
\usepackage{stfloats}
\usepackage{textcomp}
\usepackage{url}

\newcommand{\Aperta}{\Bigg( \Bigg.}
\newcommand{\Chiusa}{\Bigg. \Bigg)}

\begin{document}

\title{Local Binary and Multiclass SVMs\\ Trained on a Quantum Annealer}

\author{Enrico Zardini, Amer Delilbasic, \IEEEmembership{Student Member, IEEE}, Enrico Blanzieri, Gabriele Cavallaro, \IEEEmembership{Senior Member, IEEE}, and Davide Pastorello%
    \thanks{Enrico Zardini is with the Department of Information Engineering and Computer Science, University of Trento, Via Sommarive 9, 38123 Povo, Trento, Italy (e-mail: enrico.zardini@unitn.it).}%
    \thanks{Amer Delilbasic is with the J\"{u}lich Supercomputing Centre, Wilhelm-Johnen Stra\ss e, 52428 J\"{u}lich, Germany, also with the University of Iceland, 107 Reykjavik, Iceland, and also with ESA/ESRIN $\Phi$-lab, IT-00044 Frascati, Italy (e-mail: a.delilbasic@fz-juelich.de).}%
    \thanks{Enrico Blanzieri is with the Department of Information Engineering and Computer Science, University of Trento, Via Sommarive 9, 38123 Povo, Trento, Italy, and also with the Trento Institute for Fundamental Physics and Applications, via Sommarive 14, 38123 Povo, Trento, Italy (e-mail: enrico.blanzieri@unitn.it)}%
    \thanks{Gabriele Cavallaro is with the University of Iceland, 107 Reykjavik, Iceland, also with the J\"{u}lich Supercomputing Centre, Wilhelm-Johnen Stra\ss e, 52428 J\"{u}lich, Germany, and also with the AIDAS, 52425 J\"{u}lich, Germany (e-mail: g.cavallaro@fz-juelich.de).}%
    \thanks{Davide Pastorello is with the Department of Mathematics, Alma Mater Studiorum - Università di Bologna, Piazza di Porta San Donato 5, 40126 Bologna, Italy, and also with the Trento Institute for Fundamental Physics and Applications, via Sommarive 14, 38123 Povo, Trento, Italy (e-mail: davide.pastorello3@unibo.it).}%
    \thanks{}%
    \thanks{This work has been submitted to the IEEE for possible publication. Copyright may be transferred without notice, after which this version may no longer be accessible.}
}



\maketitle

\begin{abstract}
Support vector machines (SVMs) are widely used machine learning models (e.g., in remote sensing), with formulations for both classification and regression tasks. In the last years, with the advent of working quantum annealers, hybrid SVM models characterised by quantum training and classical execution have been introduced. These models have demonstrated comparable performance to their classical counterparts. However, they are limited in the training set size due to the restricted connectivity of the current quantum annealers. Hence, to take advantage of large datasets (like those related to Earth observation), a strategy is required. In the classical domain, local SVMs, namely, SVMs trained on the data samples selected by a $k$-nearest neighbors model, have already proven successful. Here, the local application of quantum-trained SVM models is proposed and empirically assessed. In particular, this approach allows overcoming the constraints on the training set size of the quantum-trained models while enhancing their performance. In practice, the FaLK-SVM method, designed for efficient local SVMs, has been combined with quantum-trained SVM models for binary and multiclass classification. In addition, for comparison, FaLK-SVM has been interfaced for the first time with a classical single-step multiclass SVM model (CS SVM). Concerning the empirical evaluation, D-Wave's quantum annealers and real-world datasets taken from the remote sensing domain have been employed. The results have shown the effectiveness and scalability of the proposed approach, but also its practical applicability in a real-world large-scale scenario.
\end{abstract}


\vspace{10pt}

\section{Introduction}
\label{sec:intro}
\IEEEPARstart{S}{upport} vector machines (SVMs) are supervised machine learning models designed for binary classification tasks~\cite{svms}. Specifically, an SVM aims to identify the optimal hyperplane that effectively separates data samples belonging to distinct classes. However, with the introduction of kernel functions, SVMs can go beyond linearly separable problems \cite{scholkopf_2002_kernels_learning}. Furthermore, various formulations of the learning problem exist, and also extensions to multiclass classification and regression tasks \cite{cs_svm,svr}. In the last years, with the diffusion of the quantum annealing machines produced by D-Wave \cite{D-Wave}, hybrid SVM models characterised by quantum training and classical execution have been proposed. In detail, hybrid versions for binary classification~\cite{willsch_qbsvm}, multiclass classification~\cite{delilbasic_2023_qmsvm}, and regression~\cite{pasetto_quantum_svr} tasks have been developed. These models have been evaluated mainly in the remote sensing domain (see also \cite{cavallaro_2020_qbsvm_for_rs,delilbasic_2021_qsvms_in_rs,pasetto_quantum_svr2}), showing comparable performance with respect to their classical counterparts. Nevertheless, due to the restricted connectivity of the available quantum annealers, they are limited in the training set size. Therefore, in order to leverage large datasets, a strategy is necessary.


In the classical realm, reducing the number of input samples to a machine learning model through a locality technique, such as the $k$-nearest neighbors ($k$-NN) algorithm \cite{knn}, has proven to be successful, yielding performance improvements compared to the base model. For instance, Blanzieri and Melgani have proposed and empirically assessed the $k$NNSVM classifier \cite{blanzieri_2006_local_svm}, namely, a local binary SVM trained on data samples selected by a $k$-NN model, achieving good results. Moreover, local SVMs have been theoretically characterised by researchers like Hable \cite{hable_2013_local_svm} and Meister and Steinwart \cite{meister_2016_local_svm}. However, despite the accuracy improvement and reduced training time per model (resulting from the lower number of samples employed for training), an SVM must be trained on the $k$-neighborhood of each test sample, which is a significant bottleneck in terms of execution time. To address this issue, Segata and Blanzieri have developed the Fast Local Kernel Support Vector Machine (FaLK-SVM) \cite{segata_falk_svm}, which relies on the usage of the cover tree data structure \cite{cover_trees}.

In this work, the local application of quantum-trained SVM models is proposed and empirically evaluated. Indeed, local classically-trained binary SVMs have already demonstrated to be successful, and quantum-trained SVMs have exhibited similar performance to their classical counterparts. Moreover, the usage of local quantum-trained models, as opposed to global ones, represents a valid solution to the training set size limits imposed by the connectivity of the current quantum annealers. In practice, FaLK-SVM \cite{segata_falk_svm}, the method for efficient local SVMs, has been interfaced with two quantum-trained SVM models: the quantum-trained SVM for binary classification (QBSVM) \cite{willsch_qbsvm}, and the quantum-trained SVM for multiclass classification (QMSVM) \cite{delilbasic_2023_qmsvm}. Additionally, for comparison, FaLK-SVM has been combined for the first time with CS SVM \cite{cs_svm}, the classical single-step multiclass SVM model on which QMSVM is based. Hence, the addressed tasks are binary and multiclass classification. For the empirical evaluation, D-Wave's quantum annealers and real-world datasets belonging to the remote sensing domain have been used.

The article is organized as follows: \cref{sec:background} provides some background information; \cref{sec:local-quantum-svms} presents the proposed approach and the implementation details; \cref{sec:empirical-eval} deals with the experiments performed and the results obtained; \cref{sec:conclusion} concludes the work.

\section{Background}
\label{sec:background}
This section provides some background information about quantum annealing, QUBO problems and their embedding, quantum-trained support vector machines, and local support vector machines.

\subsection{Quantum Annealing, QUBO, and Embedding}
\label{subsec:qa-qubo-and-embed}
Quantum annealing (QA) is a heuristic search used to solve optimization problems~\cite{PhysRevE.58.5355, Hauke_2020}. In particular, in QA, the solution of a given problem corresponds to the \textit{ground state} of a quantum system described by a Hamiltonian encoding the structure of the problem. In this sense, QA is related to adiabatic quantum computing, but there are some remarkable differences \cite{mcgeoch14}. Specifically, let us consider the time-dependent Hamiltonian
\begin{equation}
    H(t)=\Gamma(t) H_D+H_P,\qquad t\in[0,\tau],
\end{equation}
where $H_P$ and $H_D$ are non-commuting operators on the $n$-qubit Hilbert space $(\mathbb C^2)^{\otimes n}$ called \textit{problem Hamiltonian} and \textit{transverse field Hamiltonian}, respectively, $\Gamma$ is a positive decreasing function that attenuates the contribution of $H_D$, and $\tau$ is the evolution time. The annealing process drives the quantum system towards the ground state of $H_P$, which is designed to represent the optimization problem. 

QA can be physically realized by considering a network of qubits arranged on the vertices of a graph $(V,E)$, with $|V|=n$ and whose edges $E$ represent the couplings among the qubits. Then, the problem Hamiltonian can be defined as follows: 
\begin{equation}
    \label{HP}
    H_P=H(\boldsymbol\Theta):=\sum_{i\in V} \theta _i \sigma_z^{(i)} +\sum_{(i,j)\in E} \theta_{ij}\sigma_z^{(i)} \sigma_z^{(j)},
\end{equation}
where the real coefficients $\theta_i, \theta_{ij}$ are arranged into the matrix $\boldsymbol\Theta$ and $\sigma_z^{(i)}$ is a $2^n\times 2^n$ matrix that acts as the Pauli matrix
\begin{equation}
    \sigma_z=\left(\begin{matrix}
        1 & 0 \\
        0 & -1
    \end{matrix}\right)
\end{equation}
on the $i$-th tensor factor and as the $2\times 2$ identity matrix on the other tensor factors.
By definition, the set of eigenvalues of the problem Hamiltonian (Eq.~\ref{HP}) is the set of all possible values of the cost function given by the energy of the well-known \textit{Ising model}:
\begin{equation}
    \label{costf}
    \mathsf E(\boldsymbol{\Theta}, \boldsymbol z)=\sum_{i\in V} \theta_i z_i +\sum_{(i,j)\in E} \theta_{ij}z_i z_j,
\end{equation}
where $\boldsymbol z=(z_1,...,z_n)\in\{-1,1\}^{|V|}$.
In practice, the annealing procedure, also called \textit{cooling}, drives the system into the ground state of $H(\Theta)$, which corresponds to the spin configuration encoding the solution:
\begin{equation}
    \label{argmin_zE}
    \boldsymbol z^*=\arg\!\!\!\!\!\!\!\!\min_{\boldsymbol z\in\{-1,1\}^{|V|}} \mathsf E(\boldsymbol\Theta,\boldsymbol z).
\end{equation}
Given a problem, the annealer is initialized using a suitable choice of the weights $\boldsymbol\Theta$, and the binary variables $z_i\in\{-1,1\}$ are physically realized by the outcomes of the measurements performed on the qubits located on the vertices $V$. In order to solve a general optimization problem through QA, it is first necessary to find an \textit{encoding} of the objective function in terms of the cost function~\eqref{costf}, which is hard in general. 

However, if the quantum architecture is able to provide a fully connected graph $(V,E)$, then any Quadratic Unconstrained Binary Optimization (QUBO) problem can be directly represented into the cost function by means of the change of variables $x_i=\frac{z_i+1}{2}\in\mathbb B=\{0,1\}$. Indeed, QUBO problems are NP-hard problems of the form
\begin{equation}
    \label{eq:xtQx}
    \arg\displaystyle\min_{\!\!\!\!\!x\in\mathbb B^n} x^T Q x,
\end{equation}
where $Q$ is an upper triangular (or symmetric) matrix of real values, and they can be rewritten as
\begin{align}
    x^T Q x
        &= \displaystyle\sum_{i=1}^n q_{ii}x_{i}^2 + \displaystyle\sum_{i=1}^n\displaystyle\sum_{j=i+1}^n q_{ij} x_i x_j \nonumber \\
        &= \displaystyle\sum_{i=1}^n q_{ii}x_{i} + \displaystyle\sum_{i=1}^n\displaystyle\sum_{j=i+1}^n q_{ij} x_i x_j,
\end{align}
where $x_i^2 = x_i$ since $x_i \in \mathbb{B}$. In practice, the main diagonal of $Q$ contains the linear coefficients ($q_{ii}$), whereas the rest of the matrix contains the quadratic ones ($q_{ij}$). Although QUBO problems are unconstrained by definition, it is actually possible to introduce constraints by representing them as penalties~\cite{ayodele2022}.

In general, due to the sparseness of the available quantum annealer topologies, a direct representation of the problem is typically not possible. The solution consists in chaining together multiple physical qubits that will act as a single logical qubit. In this way, the connectivity of the annealer graph is increased at the price of reducing the number of logical qubits available and, consequently, the size of the representable problems. The mapping of the problem variables on the annealer topology is known as \textit{embedding}.

\subsection{Quantum-Trained SVM Models}
\label{subsec:qt-svms}
Quantum-trained SVMs are classical SVMs trained with quantum annealing and executed classically. In this paper, the focus is on the models for binary and multiclass classification, whose details are provided in the following.

\subsubsection{Quantum Binary SVM (QBSVM)}
\label{subsubsec:qbsvm}
In the work by Willsch et al.~\cite{willsch_qbsvm}, the standard formulation of the binary SVM has been reframed as a QUBO problem, as in \cref{eq:xtQx}. Training a binary SVM consists in the following quadratic programming problem:
\begin{gather}
    \text{minimize} \quad E=\frac{1}{2} \sum_{n m} \alpha_n \alpha_m y_n y_m k\left(\mathbf{x}_n, \mathbf{x}_m\right)-\sum_n \alpha_n \nonumber \\
    \text {subject to} \quad 0 \leq \alpha_n \leq A, \quad \sum_n \alpha_n y_n=0,
\end{gather}
for $N$ coefficients $\alpha_n\in\mathbb{R}$, where $\{\left(\mathbf{x}_n, y_n\right)\}$ is the training set of $N$ examples, $k\left(\mathbf{x}_n, \mathbf{x}_m\right)$ is the kernel function, and $A$ is the regularization parameter. The resulting classifier is defined as
\begin{equation}
    f(\mathbf{x})= \text{sign} \left( \sum_n \alpha_n y_n k\left(\mathbf{x}_n, \mathbf{x}\right)+b \right),
\end{equation}
where the bias $b$ is chosen as
\begin{equation}
    b = \frac{\sum_n \alpha_n(A - \alpha_n)\left[y_n - \sum_m \alpha_m y_m k(\mathbf{x}_n, \mathbf{x}_m)\right]}{\sum_n \alpha_n(A - \alpha_n)}.
    \label{eq:svm-b}
\end{equation}
Being already quadratic, this real-valued, constrained optimization problem can be converted to a QUBO problem by adding the constraints to the cost function as penalty terms with a multiplier $\xi$, and encoding a discretized solution space using $K$ binary variables $a_i$:
\begin{equation}
    \alpha_n=\sum_{k=0}^{K-1} B^k a_{K n+k},
\end{equation}
where $B\in{\mathbb{N}}$ is the encoding base. The corresponding QUBO problem becomes
\begin{gather}
    \text{minimize} \quad \sum_{n, m=0}^{N-1} \sum_{k, j=0}^{K-1} a_{K n+k} Q_{K n+k, K m+j} a_{K m+j}, \nonumber \\
    \begin{multlined}[b]
        Q_{K n+k, K m+j} =\frac{1}{2} B^{k+j} y_n y_m\left(k\left(\mathbf{x}_n, \mathbf{x}_m\right)+\xi\right) \\
        -\delta_{n m} \delta_{k j} B^k,
    \end{multlined}
\end{gather}
with $\delta_{ij}$ being the Kronecker delta.

Since this QUBO formulation might yield matrices not embeddable in the available quantum annealers, Willsch et al. have proposed the following approach. Firstly, the dataset is partitioned into $L$ disjoint slices. Then, for each slice, the decision functions of the $S$ best solutions (in terms of energy) obtained from the annealer are averaged. Lastly, the classifier, which corresponds to an ensemble of SVMs, is defined as
\begin{equation}
    f(\mathbf{x}) = \text{sign}\left(\frac{1}{L} \sum_{l=0}^{L-1} \left(\sum_{n=0}^{N-1} \overline{\alpha}_{n}^{(l)} y_{n}^{(l)} k(\mathbf{x}_{n}^{(l)}, \mathbf{x}) + \overline{b}^{(l)}\right)\right),
\end{equation}
where $\overline{\alpha}_{n}^{(l)}$ and $\overline{b}^{(l)}$ are the $n$-th mean coefficient and the mean bias for the $l$-th slice. Actually, averaging the $S$ best solution can be used even in scenarios where dataset splitting is not required.

\subsubsection{Quantum Multiclass SVM (QMSVM)}
\label{subsubsec:qmsvm}
As for its classical counterpart, there are two different approaches for extending QBSVM to multiclass classification. The problem can be decomposed into multiple binary problems and a QBSVM model can be trained on each subproblem, combining the obtained classifiers in an ensemble. Alternatively, a model trained to directly classify examples among multiple classes should be defined. The QMSVM approach proposed by Delilbasic et al.~\cite{delilbasic_2023_qmsvm} is based on the CS SVM model \cite{cs_svm}, which consists in the following optimization problem:
\begin{gather}
    \begin{split}
        \text{minimize} \quad E = &\frac{1}{2} \sum_{n_1,n_2=0}^{N-1} k(\mathbf{x}_{n_1},\mathbf{x}_{n_2}) \sum_{c=0}^{C-1} \tau_{n_1c}\tau_{n_2c}
        \\&- \beta \sum_{n=0}^{N-1} \sum_{c=0}^{C-1} \delta_{cy_n}\tau_{nc}
    \end{split} \nonumber \\
    \text{subject to} \quad
    \sum_{c=0}^{C-1} \tau_{nc} = 0 \enskip \forall n, \quad
    \tau_{nc} \leq 0 \enskip \forall  n, \forall c \neq y_n.
    \label{eq:multiclass_opt_n_constraints}
\end{gather}
Here, $C$ is the number of classes, $\tau_{nc} \in [-1,1]$ are the $NC$ problem variables, and $\beta$ is a regularization parameter. The resulting classifier is defined as:
\begin{equation}
    f(\mathbf{x}) = \argmax_c \sum_{n=0}^{N-1} \tau_{n c} k(\mathbf{x}_{n}, \mathbf{x}).
    \label{eq:cs-svm-decision-func}
\end{equation}
As for the binary case, $K$ binary variables are used to redefine the original optimization problem variables:
\begin{equation}
\label{eq:binary_encoding_tau}
    \tau_{nc} =-1+\frac{2}{2^{K}-1} \sum_{k=0}^{K-1} 2^k a_{nCK+cK+k}.
\end{equation}
After adding the constraints as penalty weights with a multiplier $\mu$, the QUBO matrix is defined as
\begin{multline} 
    \label{eq:msvm_qubo_matrix}
    Q_{n_1CK+c_1K+k_1,n_2CK+c_2K+k_2} =
    \\= \delta_{n_1n_2}\delta_{c_1c_2}\delta_{k_1k_2} \frac{2^{k_1+1}}{2^K-1}\Aperta -\sum_{n_3=0}^{N-1}k(\mathbf{x}_{n_1},\mathbf{x}_{n_3})
    \\- \delta_{c_1y_{n_1}}  \left(\beta+\mu\right)- 2C\mu + \mu \Chiusa
    \\+ \delta_{c_1c_2}\frac{2^{k_1+k_2+1}}{(2^K-1)^2} k(\mathbf{x}_{n_1},\mathbf{x}_{n_2}) + \delta_{n_1n_2}\frac{2^{k_1+k_2+2}\mu}{(2^K-1)^2}.
\end{multline}

To take advantage of the multiple solutions obtained from the annealer, Delilbasic et al. have proposed the following approach. Firstly, each of the $S$ best solutions (in terms of energy) is tested on a validation set (that may coincide with the training set). Subsequently, a weighted average is performed. More precisely, the weights of the solutions with an accuracy above a predefined threshold are given by a softmax function applied to the values $\mathit{multiplier} \cdot \mathit{accuracy}_s$, with $\mathit{multiplier}$ being a real value and $\mathit{accuracy}_s$ being the accuracy achieved by the $s$-th solution. Conversely, the weights of the other solutions are set to zero. The resulting mean variables $\overline{\tau}_{n c}$ are used in \cref{eq:cs-svm-decision-func} to classify the new samples. Actually, this approach allows also addressing larger datasets (part of the dataset can be used only in the weighting step).

\subsection{Local SVMs}
\label{subsec:local-svms}
Reducing the number of input samples to a classical (binary) SVM by means of a locality technique has demonstrated to be successful. In 2006, Blanzieri and Melgani have proposed and empirically evaluated the $k$NNSVM classifier \cite{blanzieri_2006_local_svm}, namely, a local SVM trained on the samples selected by a $k$-NN model, obtaining good results. Specifically, the $k$-NN and the SVM must operate in the same transformed feature space. However, for RBF kernels (like the Gaussian kernel) and polynomial kernels with degree 1, the Euclidean distance can be used as the distance metric for the $k$-NN \cite{blanzieri_2006_local_svm}. Additionally, local SVMs have been theoretically characterised by, for example, Hable \cite{hable_2013_local_svm} and Meister and Steinwart \cite{meister_2016_local_svm}. Nevertheless, despite the accuracy enhancement and the reduced training time per model (due to the lower number of samples used for training), the $k$NNSVM classifier requires to train an SVM for each test instance (unless the nearest neighbors belong to the same class), posing a serious bottleneck in terms of execution time. To address this issue, Segata and Blanzieri have devised the approach outlined below.

\subsubsection{FaLK-SVM}
\label{subsubsec:falk-svm}
FaLK-SVM \cite{segata_falk_svm} improves the execution time of the $k$NNSVM classifier \cite{blanzieri_2006_local_svm} by leveraging a data structure proposed by Beygelzimer et al. for efficient nearest-neighbor operations, i.e., the cover tree \cite{cover_trees}. Essentially, the idea consists in covering the training set with a set of local SVM models, and predicting the label of a test instance with the most suitable (pre-trained) local model. More in detail, FaLK-SVM is trained as follows: a cover tree is built on the training set; the centres of the local SVMs are selected through the cover tree, which allows the efficient retrieval of data samples that are far from one another, limiting the overlap of the local models; the local SVMs for which the local training set does not contain only one class are trained. Specifically, the selection procedure ends when each training sample belongs to the $k'$-neighborhood of at least one centre, where $k' < k$ is a hyperparameter controlling the local models redundancy. In addition, at training time, the association between each training point and the centre for which the neighbor ranking of the given training point is the smallest is determined. In this way, at prediction time, it is only necessary to identify the nearest neighbor of the test instance in the training set and execute the associated local model. Concerning the time complexity, the training step has a worst-case complexity of $O(kN \times \max(\log N, k^2))$, with $k$ being the number of nearest neighbors selected and $N$ being the number of training samples, while the prediction of a new label has a complexity of $O(\max(\log N, k))$.

In the same article, a variant of FaLK-SVM, denoted as FaLK-SVMl, has been also presented. Essentially, FaLK-SVMl incorporates a grid-search model selection procedure that is run before the training of FaLK-SVM. In practice, each combination of local model parameters is tested, using a custom $\kappa$-fold cross-validation, on $m$ local models with randomly selected centres. In this custom $\kappa$-fold cross-validation, only the $k'$ nearest neighbors of the model centre are considered for the split into folds, while the remaining $k - k'$ samples of the $k$-neighborhood are added to the training set of each $\kappa$-fold iteration. Eventually, the parameter configuration that maximizes the average accuracy of the $m$ models is selected and employed for all local models.

\section{Local Quantum-Trained SVMs}
\label{sec:local-quantum-svms}
This section introduces the proposed approach and provides details about the implementation. The code is publicly available at \url{https://github.com/ZarHenry96/local-qtrained-svms}.

\subsection{Approach}
\label{subsec:approach}
Quantum-trained support vector machines have exhibited performance akin to their classical counterparts \cite{willsch_qbsvm,delilbasic_2021_qsvms_in_rs,delilbasic_2023_qmsvm}. However, the restricted connectivity of the current quantum annealers places constraints on the size of the trainable models. Various strategies have been proposed to address this limitation and exploit larger training sets, including the construction of ensembles of SVMs \cite{willsch_qbsvm} and the weighting of the best solutions (in terms of energy) retrieved by the annealer based on their performance on a large validation set \cite{delilbasic_2023_qmsvm} (as illustrated in \cref{subsubsec:qbsvm,subsubsec:qmsvm}). The approach proposed here consists in the localised application of quantum-trained SVM models. Indeed, in the classical domain, local SVMs have demonstrated superior performance compared to their global counterparts (see \cref{subsec:local-svms} for more details). Furthermore, in this way, large training sets represent no more an issue, as each local model is trained solely on the $k$-neighborhood of the model centre.

Essentially, in this work, FaLK-SVM, the method for efficient local SVMs outlined in \cref{subsubsec:falk-svm}, has been interfaced with two quantum-trained SVM models: the quantum-trained SVM for binary classification detailed in \cref{subsubsec:qbsvm} (QBSVM), and the quantum-trained SVM for multiclass classification detailed in \cref{subsubsec:qmsvm} (QMSVM). The resulting workflow is straightforward. In fact, the only difference compared to the standard FaLK-SVM resides in the local models employed, which are trained on a quantum annealer and run classically. Actually, another innovative aspect of this study is the assessment of FaLK-SVM with local single-step multiclass classification models such as QMSVM and CS SVM, which has been taken into account for comparison (CS SVM is the basis of QMSVM, as mentioned in \cref{subsubsec:qmsvm}). Indeed, FaLK-SVM has already been assessed in a multiclass classification task, but employing a one-against-one (OAO) approach with local binary SVMs \cite{segata_2012_falk_for_rs}.

\subsection{Implementation Details}
\label{subsec:implementation}
The approach described in the previous section has been implemented building upon the FaLK-SVM implementation provided by Segata \cite{segata_falk_svm_code}. Specifically, that implementation of FaLK-SVM is written in C++, whereas the codes for QBSVM~\cite{qbsvm_code} and QMSVM~\cite{qmsvm_code} are written in Python, since Python is the only language supported by D-Wave for interacting with their quantum annealers. Therefore, to interface FaLK-SVM with the quantum-trained models, a Python class named \textit{PythonSVM} has been implemented and embedded within the FaLK-SVM C++ framework, allowing the execution of Python code within the C++ application. For this purpose, the functions, types, and macros supplied by the \textit{Python.h} header file have been employed.

From an approach-related perspective, two aspects are worth to be discussed. Firstly, to reduce the training time, the reuse of the QUBO matrix embeddings has been implemented. Basically, when a QUBO matrix of a certain size is submitted for the first time to the quantum annealer, the embedding for a complete matrix of the same size is computed, applied, and stored in memory. In all subsequent calls with QUBO matrices of that size, the precomputed embedding is retrieved and applied. This proves particularly advantageous as the QUBO matrix size is the same for all local models. Secondly, two notable features have been developed, although they have not been used in the experiments presented in this work. The first one is the local usage of the techniques illustrated in \cref{subsubsec:qbsvm,subsubsec:qmsvm} for leveraging larger datasets. This allows increasing the size of the neighborhoods used for training local models, a parameter otherwise limited by the connectivity of the annealer. The second feature pertains to multiclass classification tasks and consists in the dynamic selection of the local model based on the number of classes present in a $k$-neighborhood. Indeed, with two classes, QBSVM needs half of the binary variables compared to QMSVM.

\begin{table*}[t]
    \centering
    \caption{Local (a) and global (b) methods considered.}
    \label{tab:methods}
    \captionsetup{position=top}
    \subfloat[Local methods.\label{tab:local-methods}]{
        \centering
        \begin{tabular}{c|c|c|c}
            Name           & Classification type & Local model & Local model training \\ \hline
            FaLK-SVMl (C)  & Binary              & SVM         & Classical            \\ \hline
            FaLK-SVMl (QB) & Binary              & QBSVM       & Quantum              \\ \hline
            FaLK-SVMl (CS) & Multiclass          & CS SVM      & Classical            \\ \hline
            FaLK-SVMl (QM) & Multiclass          & QMSVM       & Quantum              \\ 
        \end{tabular}
    }
    \qquad
    \subfloat[Global methods.\label{tab:global-methods}]{
        \centering
        \begin{tabular}{c|c|c}
            Name          & Classification type & Model training \\ \hline
            SVM           & Binary              & Classical      \\ \hline
            QBSVM         & Binary              & Quantum        \\ \hline
            CS SVM        & Multiclass          & Classical      \\ \hline
            QMSVM         & Multiclass          & Quantum        \\
        \end{tabular}
    }
\end{table*}

From a model-related perspective, some modifications have been applied to the original implementations. On the FaLK-SVM front, the computation of the performance metrics and the criterion for assessing the class balance of the $m$ $k$-neighborhoods used in the local model selection procedure (of FaLK-SVMl) have been extended to multiclass classification. Regarding the grid-search local model selection, support for the parameters of the quantum-trained models has been incorporated; additionally, the number of folds $\kappa$ for the internal custom $\kappa$-fold cross-validation (10 by default) and the number of samples used to evaluate the performance of the $m$ models ($\frac{k'}{2}$ by default) have been parametrized. Eventually, a data standardization procedure has been introduced in the external canonical $\kappa$-fold cross-validation provided for assessing the performance of FaLK-SVM. On the local models front, a post-selection procedure has been implemented for the QBSVM's bias ($b$). Essentially, all values within the interval $[-10, +10]$, with a step of $0.1$, are assessed on the training set to identify the best one. Preliminary experiments have demonstrated that this approach significantly outperforms the computation of $b$ by means of \cref{eq:svm-b}. Concerning CS SVM, a C implementation of the model~\cite{cs_svm_code} has been utilized. In the local version, the CS SVM executable files are directly invoked from the Python code (after locally mapping the labels to $\{1,\dots,C\}$, if necessary). Clearly, more efficient solutions are possible. For example, in the large-scale experiment presented in \cref{sec:empirical-eval}, a custom version of CS SVM has been employed in the local setup. This more efficient version, trained with a slightly modified C code and executed via novel Python code, cannot be fully distributed due to the licensing constraints of the original CS SVM implementation~\cite{cs_svm_code}.

\section{Empirical Evaluation}
\label{sec:empirical-eval}
This section deals with the methods evaluated, the datasets employed, the experimental setup used, and the results achieved. Specifically, the classical side of the experiments has been run on a shared machine equipped with an Intel Xeon Gold 6238R processor operating at 2.20GHz and 125 GB of RAM. Instead, the quantum side has been run on the Advantage system 5.3/5.4 provided by D-Wave, a quantum annealer situated at Forschungszentrum J{\"u}lich.

\subsection{Methods}
\label{subsec:methods}
The methods taken into account in this study are reported in \cref{tab:methods}. Specifically, four local methods (\cref{tab:local-methods}) and four global methods (\cref{tab:global-methods}) have been considered here. The local ones are combinations of FaLK-SVMl (the version of FaLK-SVM with the local model selection procedure 
detailed in \cref{subsubsec:falk-svm}) and different local models: a binary and a multiclass classically-trained SVMs, namely, SVM and CS SVM, and their quantum-trained counterparts, i.e., QBSVM and QMSVM. Notice that \textit{FaLK-SVMl (C)} is the original FaLK-SVMl implementation; additional information about the other local methods are available in \cref{sec:local-quantum-svms}. Regarding the global ones, they correspond to the global application of the aforementioned classically- and quantum-trained SVM models. In particular, for the standard binary SVM, the implementation from LibSVM~\cite{libsvm} version 2.88 (the version used in the original FaLK-SVM framework) has been employed. Instead, for QBSVM and QMSVM, the strategies outlined in \cref{subsubsec:qbsvm,subsubsec:qmsvm} for handling big datasets have been utilized. Otherwise, they could have not been trained on the considered datasets, given the dataset sizes used.

\subsection{Datasets}
\label{subsec:datasets}
The methods reported in \cref{tab:methods} have been assessed on datasets taken from the remote sensing domain, a domain in which both FaLK-SVM and the quantum-trained SVMs have already shown good performance \cite{segata_2012_falk_for_rs,cavallaro_2020_qbsvm_for_rs,delilbasic_2021_qsvms_in_rs,delilbasic_2023_qmsvm}. Specifically, the datasets employed here have been generated from the SemCity Toulouse~\cite{toulouse_rs_dataset} and ISPRS Potsdam~\cite{potsdam_rs_dataset} datasets, which consist of multispectral images with multiple classes (and have been employed also in the QMSVM article~\cite{delilbasic_2023_qmsvm}). In practice, the task consists in predicting the class of each pixel. \cref{tab:datasets} provides details on the number of features and classes selected for binary and multiclass classification for both datasets. In particular, for multiclass classification, the number of classes has been restricted to three in order to maximize the number of samples that could be embedded in the annealer. Instead, the datasets sizes employed are experiment-dependent, thus they are presented in \cref{subsec:exp-setup}. Regarding the datasets generation, an equal (or approximately equal) number of samples for each class has been randomly selected from tile $4$ for Toulouse and tile $6.9$ for Potsdam, except in the large-scale experiment. Indeed, in the last experiment, the training set has been created by selecting an equal number of data points for each class from each of the $24$ Potsdam tiles labelled as training. Moreover, two distinct test sets have been generated for it: the former comprises data points randomly selected in the usual way from Potsdam tile 5.13 (a non-training tile); the latter, intended for visualization, encompasses all the data points belonging to the classes of interest within a $1000 \times 1000$ pixels square in the same tile.

\begin{table*}[t]
    \centering
    \caption{Number of features and selected classes for both basis datasets.}
    \label{tab:datasets}
    \begin{tabular}{c|c|c|c}
        Dataset name & Features number & Selected classes (binary)  & Selected classes (multiclass) \\ \hline
        Toulouse     & 8               & building, pervious surface & building, pervious surface, water                       \\ \hline
        Potsdam      & 5               & low vegetation, tree                      & building, low vegetation, tree                       \\
    \end{tabular}
\end{table*}

\begin{table*}[t]
    \centering
    \begingroup
        \renewcommand{\thefootnote}{\alph{footnote}}
        \begin{center}
            \caption{Datasets sizes for each experiment. The $\vert$ symbol in the last row separates the training set size from the (two) test sets sizes.}
            \label{tab:datasets-sizes}
            \begin{tabular}{c|c|c}
                Experiment                     & Basis datasets    & Datasets sizes                                 \\ \hline
                I - binary classification      & Toulouse, Potsdam & 500                                            \\ \hline
                II - multiclass classification & Toulouse, Potsdam & 150, 500                                       \\ \hline
                III - performance scaling      & Toulouse, Potsdam & 1000, 1500, 2000, 10000, 15000, 40000          \\ \hline
                IV - large scale (multiclass)    & Potsdam           & 11016 $\vert$ (300000, 871188\footnotemark[1]) \\
            \end{tabular}
        \end{center}
        {\small \textsuperscript{a} The occurrences of the three classes in this test set are 389900, 343438, and 137850, respectively.}
    \endgroup
\end{table*}

\subsection{Experimental Setup}
\label{subsec:exp-setup}
In this work, four experiments with different objectives have been carried out. In detail, in the first experiment, the performance of all considered binary classification methods is assessed and compared. In the second one, the same is done with the multiclass classification methods. Instead, in the third experiment, the performance scaling of the local and global fully-classical methods (both binary and multiclass) is analysed; the methods involving quantum-trained models have been omitted since the quantum annealing time consumption would have been excessive. In the final experiment, the performance of all multiclass classification methods, also the ones involving quantum-trained models, are assessed (and visualized) on a large-scale dataset. 

In the first three experiments, the performance of the methods have been evaluated using a $\kappa$-fold cross-validation procedure with ten folds ($\kappa = 10$). In practice, the input dataset is partitioned into $\kappa$ subsets, also known as folds. Then, $\kappa - 1$ folds constitute the training set, whereas the remaining one serves as the test set. This last step is iterated until each fold has been utilized once as the test set. In particular, the \textit{stratified} $\kappa$-fold cross-validation, trying to preserve the original class ratio in the folds, has been used here. In addition, to have a fair comparison, the same datasets splits have been employed for all methods. Conversely, in the final experiment, no cross-validation procedure has been used, since the input data was already divided into training set and test sets. The datasets sizes used for each of the four experiments are detailed in \cref{tab:datasets-sizes}; as already explained in \cref{subsec:datasets}, the large-scale experiment differs in the dataset generation procedure employed. Additionally, in all experiments, a data standardization procedure (involving the subtraction of the mean and the division by the standard deviation) has been applied to the training and test data features before training and running the (local/global) methods.

\begin{table*}[tb]
    \centering
    \caption{Parameters values used for binary (a) and multiclass (b) classification methods. In particular, the $m$ value between parentheses in (b) is the one used in the large-scale experiment.}
    \label{tab:methods-params}
    \captionsetup{position=top}
    \subfloat[Binary classification methods.\label{tab:bclass-methods-params}]{
        \centering
        \begin{tabular}{c|c|c|c|c|c|c|c|c|c|c|c}
            Method         & $k$ & $k'$ & $m$ & $\kappa$ (internal) & Kernel   & $\gamma$  & $A$ & $B$ & $K$ & $\xi$ & $S$ \\ \hline
            FaLK-SVMl (C)  & 80  & $60$ & 8   & 5                   & Gaussian & $-0.5, 1$ & 3   & -   & -   & -     & -   \\ \hline
            FaLK-SVMl (QB) & 80  & $60$ & 8   & 5                   & Gaussian & $-0.5, 1$ & -   & 2   & 2   & 1     & 100 \\ \hline
            SVM            & -   & -    & -   & -                   & Gaussian & 1         & 3   & -   & -   & -     & -   \\ \hline
            QBSVM          & -   & -    & -   & -                   & Gaussian & 1         & -   & 2   & 2   & 1     & 100 \\
        \end{tabular}
    } \\
    \subfloat[Multiclass classification methods.\label{tab:mclass-methods-params}]{
        \centering
        \begin{tabular}{c|c|c|c|c|c|c|c|c|c|c|c}
            Method         & $k$ & $k'$ & $m$    & $\kappa$ (internal) & Kernel   & $\gamma$  & $A$ & $K$ & $\mu$ & $\beta$ & $S$ \\ \hline
            FaLK-SVMl (CS) & 24  & $18$ & 8 (10) & 3                   & Gaussian & $-0.5, 1$ & 1   & -   & -     & -       & -   \\ \hline
            FaLK-SVMl (QM) & 24  & $18$ & 8 (10) & 3                   & Gaussian & $-0.5, 1$ & -   & 2   & 1     & 1       & 100 \\ \hline
            CS SVM         & -   & -    & -      & -                   & Gaussian & 1         & 1   & -   & -     & -       & -   \\ \hline
            QMSVM          & -   & -    & -      & -                   & Gaussian & 1         & -   & 2   & 1     & 1       & 100 \\
        \end{tabular}
    }
\end{table*}

Regarding the parameters values employed for the various methods, they are detailed in \cref{tab:methods-params}. Let us consider first the binary classification methods (\cref{tab:bclass-methods-params}). The training neighborhood size ($k$) for the local methods has been set to 80, a value close to the maximum number of samples that can be embedded in the present quantum annealers with the QBSVM QUBO formulation (considering the values of $B$ and $K$, and finding the embedding for a complete matrix). In addition, a relatively-high degree of local models overlap (regulated by $k'$) has been utilized. Concerning FaLK-SVMl's local model selection, 8 local models ($m$) and 5 folds ($\kappa$ \textit{internal}) have been used; additionally, all $k'$ samples have been employed in the assessment of the $m$ local models. Specifically, the grid search has been applied only to the Gaussian kernel width $\gamma$, to find the best value between $-0.5$ and $1$, with $-0.5$ corresponding to the usage of the median of the distances in the neighborhood as the kernel width. Therefore, with $\gamma = -0.5$, each local SVM model could have a different kernel width value (more details can be found in the FaLK-SVM article~\cite{segata_falk_svm}). Conversely, for the global methods, $\gamma$ has been fixed to $1$ (as their implementations do not support the usage of the median as the kernel width). To ensure a fair comparison between classically- and quantum-trained models, the SVM cost parameter ($A$) has been set to 3 (for QBSVM, $A$ is determined by $B$ and $K$). Concerning the QBSVM-specific parameters, the encoding basis ($B$) and the number of binary variables per coefficient ($K$) have been set to small values, to enable the embedding of an adequate number of training samples. Furthermore, the penalty coefficient ($\xi$) has been set to 1 (the same value employed for QMSVM, where it is denoted as $\mu$), and the best 100 solutions found by the annealer have been taken into account for averaging ($S$). Lastly, for the global application of QBSVM, a stratified training data split has been used, with each slice (except the last one) having a number of samples equal to $k$.

Similar considerations apply to the multiclass classification methods (\cref{tab:mclass-methods-params}). Indeed, the training neighborhood size ($k$) for the local methods has been set to 24, a value close to the maximum number of samples that can be embedded in the current quantum annealers with the QMSVM QUBO formulation (considering the values of $C = 3$ and $K$, and finding the embedding for a complete matrix). Concerning the local model selection, 3 folds ($\kappa$ \textit{internal}) have been utilized, due the smaller number of samples involved. Moreover, in the large-scale experiment, 10 local models ($m$) have been used instead of 8. Instead, the CS SVM cost parameter ($A$) has been set to 1 for a fair comparison with the QMSVM-based methods. In fact, the following relationship holds: $A = 1 / \beta$. Regarding the QMSVM-related parameters ($K, \mu, \beta, S$), the same configuration employed in the QMSVM article~\cite{delilbasic_2023_qmsvm} (where $K$ is denoted as $B$) has been used here. The accuracy threshold definition ($thr = 0.2 * \min(acc) + 0.8 * \max(acc)$) and the $\mathit{multiplier}$ value (10) for weighting the $S$ best solutions returned by the annealer (on the local $k$-neighborhood) have also been adopted from that work. In contrast, the $max\_min\_ratio$ used to prune the small QUBO matrix coefficients has been set to a high value ($1000$), rendering the pruning procedure ineffective (the embedding is computed for a complete matrix here). Eventually, for the global application of QMSVM, a stratified random selection of the training samples has been used, with a number of chosen samples equal to $k$.

The quantum annealing parameters employed in the experiments are detailed in \cref{tab:annealing-params}. Specifically, this is the same configuration used in the QMSVM article~\cite{delilbasic_2023_qmsvm}. With the setup employed in this article, training a single QBSVM model requires approximately $0.360s$ of quantum annealing time (the number of binary variables involved is 160). A slightly shorter time interval is necessary for a QMSVM model (for which the number of binary variables is 144).

\begin{table}[htb]
    \centering
    \caption{Quantum annealing parameters.}
    \label{tab:annealing-params}
    \begin{tabular}{c|c|c}
         Number of reads & Annealing time & Chain strength  \\ \hline
         1000            & 200$\mu s$     & 1               \\
    \end{tabular}
\end{table}

\subsection{Results}
\label{subsec:results}
The performance metric chosen for the methods evaluation is the classification accuracy, which is given by
\begin{equation}
    \label{eq:accuracy}
    \mathit{accuracy} = \frac{\mathit{number}\ \mathit{of}\ \mathit{correctly}\ \mathit{classified}\ \mathit{samples}}{\mathit{total}\ \mathit{number}\ \mathit{of}\ \mathit{samples}}.
\end{equation}
In particular, in the first three experiments, employing the $\kappa$-fold cross-validation, the accuracy calculated on the entire dataset (considering the predictions from the $\kappa$ models) is reported. Given that a stratified $\kappa$-fold cross-validation has been used, the folds may not have precisely the same number of elements. Consequently, there could be a small discrepancy between the reported accuracy and the average accuracy over folds. Nevertheless, this difference is negligible. Conversely, in the last experiment, the accuracy achieved on the two test sets is presented. Moreover, for the second test set, which is not (class-) balanced, two additional metrics are reported. These metrics are the balanced accuracy~\cite{balanced_accuracy}, corresponding to the average recall over classes, and the F1 score~\cite{f1_score} (namely, the harmonic mean of precision and recall) averaged over classes.

\subsubsection{Binary Classification}
\label{subsubsec:binary-classification}
In the first experiment, the performance of the binary classification methods have been assessed. The results achieved are reported in \cref{tab:binary-acc}. In practice, in the case of Toulouse, all methods have obtained good results, but the entirely-classical methods have demonstrated superior performance overall, and the local methods have outperformed their global counterparts. Conversely, in the case of Potsdam, the methods have obtained worse results overall, with the classical SVM achieving the best performance, and FaLK-SVMl (QB) outperforming its classical counterpart (albeit not by much). Concerning QBSVM, it has shown the worst performance among the evaluated methods also in this case. Therefore, the local application of QBSVM has proven effective. In fact, it has achieved results not too far from, if not better than, its classical counterpart.

\begin{table*}[tb]
    \centering
    \caption{Accuracy achieved by binary classification methods (columns) on different datasets (rows) of size 500. For FaLK-SVMl, the average number of local models (over folds) and the size of the local models are reported between square brackets; instead, for QBSVM, the number of models (all with size 80, except the last one with size 50) is shown between square brackets.}
    \label{tab:binary-acc}
    \begin{tabular}{c|c|c|c|c}
                       & FaLK-SVMl (C)     & FaLK-SVMl (QB)    & SVM    & QBSVM          \\ \hline
        Toulouse (500) & 92.4\% [15.9, 80] & 89.8\% [15.9, 80] & 91.8\% & 88.2\% [6, ] \\ \hline
        Potsdam  (500) & 69.8\% [16.5, 80] & 70.4\% [16.5, 80] & 73.0\% & 68.6\% [6, ] \\ 
    \end{tabular}
\end{table*}

\subsubsection{Multiclass Classification}
\label{subsubsec:multiclass-classification}
In the second experiment, the performance of the multiclass classification methods have been assessed. The results are reported in \cref{tab:multiclass-acc}. Let us focus first on the smaller datasets (size 150), for which the average number of local models aligns with that of the binary classification methods in the first experiment. Specifically, in the case of Toulouse, the local methods have obtained the best results and the entirely-classical ones (both local and global) have outperformed their quantum-trained counterparts. The overall results obtained are good. Instead, in the case of Potsdam, the accuracy values are lower, yet the trend is similar. The exception is represented by FaLK-SVMl (QM), which has performed worse than not only FaLK-SVMl (CS) but also CS SVM. Nevertheless, with larger datasets (size 500), FaLK-SVMl (QM) has been the top-performing method, surpassing both FaLK-SVMl (CS) and CS SVM. Regarding the performance-based ordering of the other methods, it is the same. Overall, the larger dataset size has proven advantageous, particularly in the case of Toulouse. Eventually, even in this experiment, the global quantum-trained model (QMSVM) has exhibited the worst performance among the methods evaluated. In summary, this second experiment has proven the efficacy of locally applying both classically- and quantum-trained single-step multiclass SVMs, with the quantum-trained ones being slightly better in the case of larger datasets.

\begin{table*}[tb]
    \centering
    \caption{Accuracy achieved by multiclass classification methods (columns) on datasets (rows) of two different sizes. For FaLK-SVMl, the average number of local models (over folds) and the size of the local models are reported between square brackets; instead, for QMSVM, the size of the model is shown between square brackets.}
    \label{tab:multiclass-acc}
    \begin{tabular}{c|c|c|c|c}
                       & FaLK-SVMl (CS)    & FaLK-SVMl (QM)    & CS SVM & QMSVM          \\ \hline
        Toulouse (150) & 79.3\% [14.2, 24] & 77.3\% [14.2, 24] & 76.0\% & 72.0\% [, 24]  \\ \hline
        Potsdam  (150) & 72.0\% [14.3, 24] & 66.0\% [14.3, 24] & 70.0\% & 60.7\% [, 24]  \\ \hline
        Toulouse (500) & 85.2\% [50.2, 24] & 85.4\% [50.2, 24] & 80.8\% & 73.0\% [, 24]  \\ \hline
        Potsdam  (500) & 72.6\% [47.3, 24] & 72.8\% [47.3, 24] & 71.4\% & 58.8\% [, 24]  \\ 
    \end{tabular}
\end{table*}

\subsubsection{Performance Scaling (Classical Methods)}
\label{subsubsec:performance-scaling}
In the third experiment, a performance scaling analysis has been carried out on the classical (binary and multiclass) local methods, taking into account their global counterparts for comparison. In fact, in the previous experiments, FaLK-SVMl (QB) and FaLK-SVMl (QM) have obtained results not too far (except in one instance) from their classical counterparts, which can then be used as indicators of performance. Moreover, the quantum annealing time consumption would have been excessive for the available resources.

\begin{table*}[tb]
    \centering
    \caption{Accuracy achieved by classical binary classification methods (columns) on datasets with different sizes (rows). For comparison, the pertinent results presented in \cref{tab:binary-acc} are reported here. In particular, for FaLK-SVMl (C), the average number of local models (over folds) and the size of the local models are reported between square brackets.}
    \label{tab:bclass-perf-scaling}
    \captionsetup{position=top}
    \subfloat[Toulouse (binary).\label{tab:toulouse-b-perf-scaling}]{
        \centering
        \begin{tabular}{c|c|c}
                  & FaLK-SVMl (C)       & SVM    \\ \hline
            500   & 92.4\% [15.9, 80]   & 91.8\% \\ \hline
            1000  & 92.1\% [33.7, 80]   & 91.9\% \\ \hline
            1500  & 91.9\% [54.4, 80]   & 92.4\% \\ \hline
            2000  & 91.5\% [73.8, 80]   & 91.7\% \\ \hline
            10000 & 91.6\% [423.1, 80]  & 92.2\% \\ \hline
            15000 & 91.5\% [645.3, 80]  & 92.0\% \\ \hline
            40000 & 91.7\% [1788.9, 80] & 92.1\% \\
        \end{tabular}
    }
    \qquad
    \subfloat[Potsdam (binary).\label{tab:potsdam-b-perf-scaling}]{
        \centering
        \begin{tabular}{c|c|c}
                  & FaLK-SVMl (C)       & SVM    \\ \hline
            500   & 69.8\% [16.5, 80]   & 73.0\% \\ \hline
            1000  & 73.2\% [34.6, 80]   & 71.8\% \\ \hline
            1500  & 72.3\% [54.6, 80]   & 72.3\% \\ \hline
            2000  & 71.5\% [74.6, 80]   & 72.6\% \\ \hline
            10000 & 74.5\% [400.2, 80]  & 74.7\% \\ \hline
            15000 & 74.1\% [603.4, 80]  & 75.0\% \\ \hline
            40000 & 74.8\% [1641.9, 80] & 75.5\% \\
        \end{tabular}
    }
\end{table*}

\begin{table*}[tb]
    \centering
    \caption{Accuracy achieved by classical multiclass classification methods (columns) on datasets with different sizes (rows). For comparison, the pertinent results presented in \cref{tab:multiclass-acc} are reported here. In particular, for FaLK-SVMl (CS), the average number of local models (over folds) and the size of the local models are reported between square brackets.}
    \label{tab:multiclass-perf-scaling}
    \captionsetup{position=top}
    \subfloat[Toulouse (multiclass).\label{tab:toulouse-m-perf-scaling}]{
        \centering
        \begin{tabular}{c|c|c}
                  & FaLK-SVMl (CS)      & CS SVM \\ \hline
            150   & 79.3\% [14.2, 24]   & 76\%   \\ \hline
            500   & 85.2\% [50.2, 24]   & 80.8\% \\ \hline
            1000  & 87.3\% [104.7, 24]  & 81.7\% \\ \hline
            1500  & 88.1\% [165.9, 24]  & 82.1\% \\ \hline
            2000  & 86.4\% [221.5, 24]  & 81.4\% \\ \hline
            10000 & 87.8\% [1167.8, 24] & 81.2\% \\ \hline
            15000 & 88.2\% [1761.7, 24] & 81.2\% \\ \hline
            40000 & 89.1\% [4779.5, 24] & 81.3\% \\
        \end{tabular}
    }
    \qquad
    \subfloat[Potsdam (multiclass).\label{tab:potsdam-m-perf-scaling}]{
        \centering
        \begin{tabular}{c|c|c}
                  & FaLK-SVMl (CS)      & CS SVM \\ \hline
            150   & 72.0\% [14.3, 24]   & 70\%   \\ \hline
            500   & 72.6\% [47.3, 24]   & 71.4\% \\ \hline
            1000  & 72.8\% [100.8, 24]  & 69.2\% \\ \hline
            1500  & 73.1\% [155.5, 24]  & 70.1\% \\ \hline
            2000  & 74.0\% [211.5, 24]  & 70.4\% \\ \hline
            10000 & 77.7\% [1079.9, 24] & 69.8\% \\ \hline
            15000 & 77.9\% [1646.0, 24] & 69.6\% \\ \hline
            40000 & 79.0\% [4465.1, 24] & 69.4\% \\
        \end{tabular}
    }
\end{table*}

The results are showcased in \cref{tab:bclass-perf-scaling,tab:multiclass-perf-scaling}. Let us consider binary classification (\cref{tab:bclass-perf-scaling}) first. In the case of Toulouse, the performance of both FaLK-SVMl (C) and SVM have been quite stable while increasing the dataset size, with little variations (worsening and improvement, respectively) compared to the baseline size of 500. However, the accuracy values for dataset size 500 were already really high. In contrast, in the case of Potsdam, where the initial performance were worse, an improvement has been observed for both FaLK-SVMl (C) and SVM. Specifically, the improvement has been more marked yet less consistent for FaLK-SVMl (C), and less marked but more consistent (after an initial drop) for SVM. Regarding multiclass classification (\cref{tab:multiclass-perf-scaling}), the scenario is the following: in both cases (Toulouse and Potsdam), the performance of FaLK-SVMl (CS) have almost always improved while increasing the dataset size; conversely, the performance of CS SVM have either significantly improved at the beginning and then remained quite stable (Toulouse) or fluctuated around the initial value (Potsdam). In addition, despite the slight drop for sizes 2000 and 10000, the enhancement for FaLK-SVMl (CS) has been more marked for Toulouse. Essentially, this experiment has proven that local methods can leverage larger datasets, particularly FaLK-SVMl (CS), which has outperformed CS SVM in all tests. In contrast, FaLK-SVMl (C) has been outperformed by SVM in almost all cases (albeit not by much), but has shown good stability when its performance have not improved (Toulouse).

\subsubsection{Large Scale (Multiclass)}
\label{subsubsec:large-scale}
In the last experiment, the performance of all multiclass classification methods have been assessed (without $\kappa$-fold cross-validation) on a large-scale dataset based on Potsdam, featuring one training set and two test sets (created as explained in \cref{subsec:datasets}). The objective is to showcase the performance achievable by locally applying quantum-trained SVM models in a large-scale real-world scenario. Due to the restricted quantum annealing resources available, only multiclass classification and Potsdam have been taken into account.

\begin{table*}[tb]
    \centering
    \caption{Accuracy achieved by multiclass classification methods (columns) on a large-scale dataset based on Potsdam, characterised by 11016 training samples and 300000 test samples. For FaLK-SVMl, the number of local models and the size of the local models are reported between square brackets; instead, for QMSVM, the size of the model is shown between square brackets.}
    \label{tab:large-scale-acc}
    \begin{tabular}{c|c|c|c|c}
                  & FaLK-SVMl (CS)    & FaLK-SVMl (QM)    & CS SVM & QMSVM         \\ \hline
        Accuracy  & 74.2\% [1326, 24] & 73.8\% [1326, 24] & 69.9\% & 55.6\% [, 24] \\
    \end{tabular}
\end{table*}

\begin{figure*}[tb]
    \centering
    \subfloat[Original (RGB).\label{fig:large-scale-vis-original}]{
        \centering
        \includegraphics[width=0.320\linewidth]{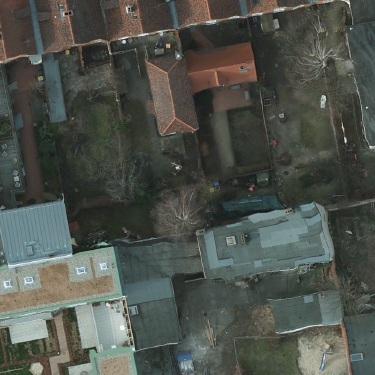}
    }
    \subfloat[Ground truth.\label{fig:large-scale-vis-ground-truth}]{
        \centering
        \includegraphics[width=0.320\linewidth]{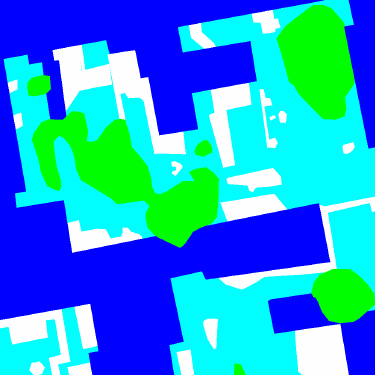}
    }
    \subfloat[FaLK-SVMl (CS).\label{fig:large-scale-vis-falk-svm-cs}]{
        \centering
        \includegraphics[width=0.320\linewidth]{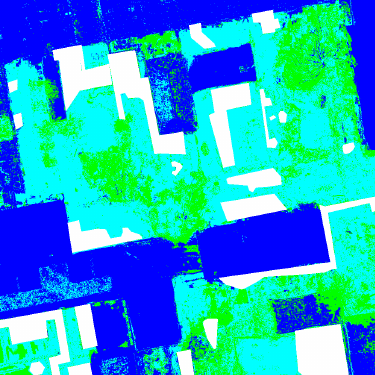}
    } \\ \vspace{5pt}
    \subfloat[FaLK-SVMl (QM).\label{fig:large-scale-vis-falk-svm-qm}]{
        \centering
        \includegraphics[width=0.320\linewidth]{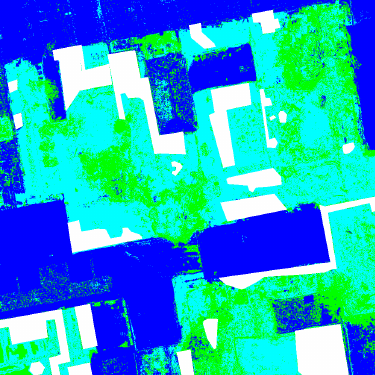}
    }
    \subfloat[CS SVM.\label{fig:large-scale-vis-cs-svm}]{
        \centering
        \includegraphics[width=0.320\linewidth]{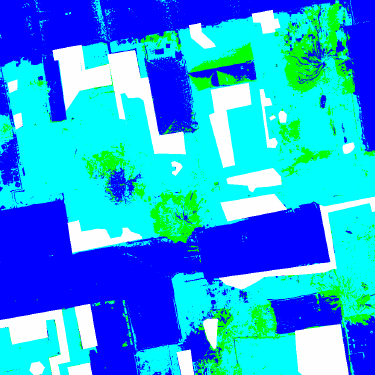}
    }
    \subfloat[QMSVM.\label{fig:large-scale-vis-qm-svm}]{
        \centering
        \includegraphics[width=0.320\linewidth]{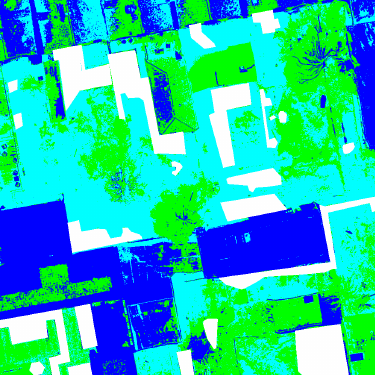}
    }
    \caption{Visualization of the results obtained by the multiclass classification methods on the second test set for the large-scale experiment. The corresponding performance metrics are provided in \cref{tab:large-scale-visualization-acc}. Color legend: blue = building, light blue = low vegetation, green = tree.}
    \label{fig:large-scale-visualization}
\end{figure*}

The results achieved on the first test set are presented in \cref{tab:large-scale-acc}. Specifically, the local methods have outperformed the global ones, and the entirely-classical methods have demonstrated superior performance compared to their quantum-trained counterparts. This last point seems to contradict the observations made in \cref{subsubsec:multiclass-classification} about the local methods, with FaLK-SVMl (QM) performing better than FaLK-SVMl (CS) on larger datasets. Nevertheless, the performance differences are relatively small. Furthermore, in this case, no $\kappa$-fold cross-validation has been utilized. In fact, the training set has been constructed by randomly sampling data points from various tiles, and the test data points have been randomly chosen from a different tile. Therefore, the task is somehow different. Despite this aspect, these first results align well with the expectations based on the outcomes of the previous experiments. Actually, a larger $k$ value (100, with $k'=75$) has also been evaluated for FaLK-SVM (CS) in this same setup. The performance obtained were slightly worse (accuracy = 73.6\%), but CS SVM has already shown that it does not really take advantage of larger training sets, particularly in the case of Potsdam (see \cref{subsubsec:performance-scaling}). Regarding the second test set, the results are presented in \cref{tab:large-scale-visualization-acc}. Unexpectedly, CS SVM has obtained the highest accuracy on this second test set, performing better than both local methods. However, given the unbalanced nature of this test set, different performance metrics should be taken into account. Here, the balanced accuracy and the average F1 score (over classes) have been considered (their values are reported in the same table). In detail, according to these metrics, both FaLK-SVMl (CS) and FaLK-SVMl (QM) have outperformed CS SVM, aligning with the trend observed for the first test set and in the previous experiments. This phenomenon is caused by CS SVM's tendency to predict more frequently the two most represented classes (building and low vegetation) in the test set, misclassifying the less common one (tree). This behaviour can be easily noticed in \cref{fig:large-scale-visualization}, where the predictions of the different methods are shown. In conclusion, this final experiment has proven the practical applicability of local quantum-trained SVMs (in particular, the multiclass one) in a large-scale scenario. In fact, the results obtained by FaLK-SVMl (QM) are quite good and comparable to those achieved by its classical counterpart.

\begin{table*}[tb]
    \centering
    \caption{Accuracy, balanced accuracy, and average F1 score (over classes) achieved by multiclass classification methods (columns) on a second large-scale test set based on Potsdam and consisting of 871188 samples (389900, 343438, 137850). These results have been obtained using the trained models employed for \cref{tab:large-scale-acc}. For FaLK-SVMl, the number of local models and the size of the local models are reported between square brackets; instead, for QMSVM, the size of the model is shown between square brackets.}
    \label{tab:large-scale-visualization-acc}
    \begin{tabular}{c|c|c|c|c}
                          & FaLK-SVMl (CS)    & FaLK-SVMl (QM)    & CS SVM & QMSVM         \\ \hline
        Accuracy          & 77.7\% [1326, 24] & 76.6\% [1326, 24] & 78.6\% & 62.0\% [, 24] \\ \hline
        Balanced accuracy & 72.4\% [1326, 24] & 71.7\% [1326, 24] & 68.5\% & 61.6\% [, 24] \\ \hline
        Average F1 score  & 71.9\% [1326, 24] & 70.9\% [1326, 24] & 69.0\% & 59.2\% [, 24] \\
    \end{tabular}
\end{table*}

\section{Conclusion}
\label{sec:conclusion}
In this article, the local application of quantum-trained SVM models, with the aim of enabling their usage on large datasets and enhancing their performance, has been introduced and empirically assessed in the remote sensing domain. Specifically, here, a method for efficient local SVMs (FaLK-SVM) has been interfaced with two quantum-trained SVM models: an SVM model for binary classification (QBSVM) and an SVM model for multiclass classification (QMSVM). In addition, for comparison, FaLK-SVM has been paired for the first time with a classical single-step multiclass classification model (CS SVM). Details about the implementation, such as the post-selection procedure for QBSVM's bias, and the experimental setup have been provided. The results have demonstrated the effectiveness of the approach, with the local applications of QBSVM and QMSVM obtaining results not too far from, if not better than, their classical counterparts. The local application of CS SVM has also yielded good results, consistently outperforming its global counterpart. Furthermore, the performance scaling analysis carried out on the classical local methods, serving as performance indicators for the quantum-trained ones, has revealed their ability to take advantage of larger datasets. Ultimately, the last experiment has proven the practical applicability of the local quantum-trained SVMs (specifically, the multiclass one) in a real-world large-scale scenario.

Future work includes the assessment of these local quantum-trained methods on datasets taken from a different domain, using different parameter configurations and a higher number of reads. Another interesting possibility consists in the development of a local version of the quantum-trained support vector regression model~\cite{pasetto_quantum_svr} (not considered here).

\section*{Acknowledgments}
This work was supported by Q@TN, the joint lab between University of Trento, FBK-Fondazione Bruno Kessler, INFN-National Institute for Nuclear Physics and CNR-National Research Council. In addition, this work was partially supported by project SERICS (PE00000014) under the MUR National Recovery and Resilience Plan funded by the European Union - NextGenerationEU. The authors gratefully acknowledge the J\"ulich Supercomputing Center (\url{https://www.fz-juelich.de/ias/jsc}) for funding this project by providing computing time on the D-Wave Advantage\texttrademark{} System JUPSI through the J\"ulich UNified Infrastructure for Quantum computing (JUNIQ). Lastly, the authors gratefully acknowledge the Italian Ministry of University and Research (MUR), which, under the initiative "Dipartimenti di Eccellenza 2018-2022 (Legge 232/2016)", has provided the (classical) computational resources used in the experiments.

\bibliographystyle{IEEEtran}
\bibliography{bibliography}

\begin{thebibliography}{10}
\providecommand{\url}[1]{#1}
\csname url@samestyle\endcsname
\providecommand{\newblock}{\relax}
\providecommand{\bibinfo}[2]{#2}
\providecommand{\BIBentrySTDinterwordspacing}{\spaceskip=0pt\relax}
\providecommand{\BIBentryALTinterwordstretchfactor}{4}
\providecommand{\BIBentryALTinterwordspacing}{\spaceskip=\fontdimen2\font plus
\BIBentryALTinterwordstretchfactor\fontdimen3\font minus \fontdimen4\font\relax}
\providecommand{\BIBforeignlanguage}[2]{{%
\expandafter\ifx\csname l@#1\endcsname\relax
\typeout{** WARNING: IEEEtran.bst: No hyphenation pattern has been}%
\typeout{** loaded for the language `#1'. Using the pattern for}%
\typeout{** the default language instead.}%
\else
\language=\csname l@#1\endcsname
\fi
#2}}
\providecommand{\BIBdecl}{\relax}
\BIBdecl

\bibitem{svms}
N.~Cristianini and J.~Shawe-Taylor, \emph{{An Introduction to Support Vector Machines and Other Kernel-based Learning Methods}}.\hskip 1em plus 0.5em minus 0.4em\relax Cambridge University Press, 2000.

\bibitem{scholkopf_2002_kernels_learning}
B.~Sch{\"o}lkopf and A.~J. Smola, \emph{{Learning with kernels: support vector machines, regularization, optimization, and beyond}}.\hskip 1em plus 0.5em minus 0.4em\relax MIT press, 2002.

\bibitem{cs_svm}
K.~Crammer and Y.~Singer, ``{On the Algorithmic Implementation of Multiclass Kernel-Based Vector Machines},'' \emph{Journal of Machine Learning Research}, vol.~2, p. 265–292, 3 2002.

\bibitem{svr}
\BIBentryALTinterwordspacing
H.~Drucker, C.~J.~C. Burges, L.~Kaufman, A.~Smola, and V.~Vapnik, ``{Support Vector Regression Machines},'' in \emph{Advances in Neural Information Processing Systems}, M.~Mozer, M.~Jordan, and T.~Petsche, Eds., vol.~9.\hskip 1em plus 0.5em minus 0.4em\relax MIT Press, 1996. [Online]. Available: \url{https://proceedings.neurips.cc/paper_files/paper/1996/file/d38901788c533e8286cb6400b40b386d-Paper.pdf}
\BIBentrySTDinterwordspacing

\bibitem{D-Wave}
{D-Wave Systems Inc.}, ``{D-Wave Systems},'' \url{https://www.dwavesys.com}, 2023, last access on 16 Jan 2024.

\bibitem{willsch_qbsvm}
\BIBentryALTinterwordspacing
D.~Willsch, M.~Willsch, H.~{De Raedt}, and K.~Michielsen, ``{Support vector machines on the D-Wave quantum annealer},'' \emph{Computer Physics Communications}, vol. 248, p. 107006, 2020. [Online]. Available: \url{https://www.sciencedirect.com/science/article/pii/S001046551930342X}
\BIBentrySTDinterwordspacing

\bibitem{delilbasic_2023_qmsvm}
A.~Delilbasic, B.~Le~Saux, M.~Riedel, K.~Michielsen, and G.~Cavallaro, ``{A Single-Step Multiclass SVM Based on Quantum Annealing for Remote Sensing Data Classification},'' \emph{IEEE Journal of Selected Topics in Applied Earth Observations and Remote Sensing}, pp. 1--12, 2023.

\bibitem{pasetto_quantum_svr}
E.~Pasetto, A.~Delilbasic, G.~Cavallaro, M.~Willsch, F.~Melgani, M.~Riedel, and K.~Michielsen, ``{Quantum Support Vector Regression for Biophysical Variable Estimation in Remote Sensing},'' in \emph{IGARSS 2022 - 2022 IEEE International Geoscience and Remote Sensing Symposium}, 2022, pp. 4903--4906.

\bibitem{cavallaro_2020_qbsvm_for_rs}
G.~Cavallaro, D.~Willsch, M.~Willsch, K.~Michielsen, and M.~Riedel, ``{Approaching Remote Sensing Image Classification with Ensembles of Support Vector Machines on the D-Wave Quantum Annealer},'' in \emph{IGARSS 2020 - 2020 IEEE International Geoscience and Remote Sensing Symposium}, 2020, pp. 1973--1976.

\bibitem{delilbasic_2021_qsvms_in_rs}
A.~Delilbasic, G.~Cavallaro, M.~Willsch, F.~Melgani, M.~Riedel, and K.~Michielsen, ``{Quantum Support Vector Machine Algorithms for Remote Sensing Data Classification},'' in \emph{2021 IEEE International Geoscience and Remote Sensing Symposium IGARSS}, 2021, pp. 2608--2611.

\bibitem{pasetto_quantum_svr2}
E.~Pasetto, M.~Riedel, F.~Melgani, K.~Michielsen, and G.~Cavallaro, ``{Quantum SVR for Chlorophyll Concentration Estimation in Water With Remote Sensing},'' \emph{IEEE Geoscience and Remote Sensing Letters}, vol.~19, pp. 1--5, 2022.

\bibitem{knn}
E.~Fix and J.~L. Hodges, ``{Discriminatory Analysis, Nonparametric Discrimination: Consistency Properties},'' USAF School of Aviation Medicine, Randolph Field, Tech. Rep.~4, 1951.

\bibitem{blanzieri_2006_local_svm}
E.~Blanzieri and F.~Melgani, ``{An Adaptive SVM Nearest Neighbor Classifier for Remotely Sensed Imagery},'' in \emph{2006 IEEE International Symposium on Geoscience and Remote Sensing}, 2006, pp. 3931--3934.

\bibitem{hable_2013_local_svm}
\BIBentryALTinterwordspacing
R.~Hable, ``{Universal Consistency of Localized Versions of Regularized Kernel Methods},'' \emph{Journal of Machine Learning Research}, vol.~14, no.~5, pp. 153--186, 2013. [Online]. Available: \url{http://jmlr.org/papers/v14/hable13a.html}
\BIBentrySTDinterwordspacing

\bibitem{meister_2016_local_svm}
\BIBentryALTinterwordspacing
M.~Meister and I.~Steinwart, ``{Optimal Learning Rates for Localized SVMs},'' \emph{Journal of Machine Learning Research}, vol.~17, no. 194, pp. 1--44, 2016. [Online]. Available: \url{http://jmlr.org/papers/v17/14-023.html}
\BIBentrySTDinterwordspacing

\bibitem{segata_falk_svm}
\BIBentryALTinterwordspacing
N.~Segata and E.~Blanzieri, ``{Fast and Scalable Local Kernel Machines},'' \emph{Journal of Machine Learning Research}, vol.~11, no.~64, pp. 1883--1926, 2010. [Online]. Available: \url{http://jmlr.org/papers/v11/segata10a.html}
\BIBentrySTDinterwordspacing

\bibitem{cover_trees}
\BIBentryALTinterwordspacing
A.~Beygelzimer, S.~Kakade, and J.~Langford, ``{Cover Trees for Nearest Neighbor},'' in \emph{Proceedings of the 23rd International Conference on Machine Learning}, ser. ICML '06.\hskip 1em plus 0.5em minus 0.4em\relax New York, NY, USA: Association for Computing Machinery, 2006, p. 97–104. [Online]. Available: \url{https://doi.org/10.1145/1143844.1143857}
\BIBentrySTDinterwordspacing

\bibitem{PhysRevE.58.5355}
\BIBentryALTinterwordspacing
T.~Kadowaki and H.~Nishimori, ``Quantum annealing in the transverse ising model,'' \emph{Phys. Rev. E}, vol.~58, pp. 5355--5363, Nov 1998. [Online]. Available: \url{https://link.aps.org/doi/10.1103/PhysRevE.58.5355}
\BIBentrySTDinterwordspacing

\bibitem{Hauke_2020}
\BIBentryALTinterwordspacing
P.~Hauke, H.~G. Katzgraber, W.~Lechner, H.~Nishimori, and W.~D. Oliver, ``Perspectives of quantum annealing: methods and implementations,'' \emph{Reports on Progress in Physics}, vol.~83, no.~5, p. 054401, may 2020. [Online]. Available: \url{https://dx.doi.org/10.1088/1361-6633/ab85b8}
\BIBentrySTDinterwordspacing

\bibitem{mcgeoch14}
C.~C.~McGeoch, \emph{{Adiabatic Quantum Computation and Quantum Annealing}}.\hskip 1em plus 0.5em minus 0.4em\relax Springer Cham, 2014.

\bibitem{ayodele2022}
M.~Ayodele, ``Penalty weights in qubo formulations: Permutation problems,'' in \emph{Evolutionary Computation in Combinatorial Optimization}, L.~P{\'e}rez~C{\'a}ceres and S.~Verel, Eds.\hskip 1em plus 0.5em minus 0.4em\relax Cham: Springer International Publishing, 2022, pp. 159--174.

\bibitem{segata_2012_falk_for_rs}
\BIBentryALTinterwordspacing
N.~Segata, E.~Pasolli, F.~Melgani, and E.~Blanzieri, ``{Local SVM approaches for fast and accurate classification of remote-sensing images},'' \emph{International Journal of Remote Sensing}, vol.~33, no.~19, pp. 6186--6201, 2012. [Online]. Available: \url{https://doi.org/10.1080/01431161.2012.678947}
\BIBentrySTDinterwordspacing

\bibitem{segata_falk_svm_code}
N.~Segata, ``{FaLKM-lib v1.0: a Library for Fast Local Kernel Machines},'' DISI, University of Trento, Italy, Tech. Rep. DISI-09-025, 2009, software available at \url{http://disi.unitn.it/~segata/FaLKM-lib}.

\bibitem{qbsvm_code}
D.~Willsch, G.~Cavallaro, and M.~Willsch, ``{QA SVM implementation},'' \url{https://gitlab.jsc.fz-juelich.de/sdlrs/quantum-svm-algorithms-for-rs-data-classification/-/tree/master/experiments/QA_SVM?ref_type=heads}, 2021, last access on 15 Dec 2023.

\bibitem{qmsvm_code}
A.~Delilbasic, ``{QMSVM implementation},'' \url{https://gitlab.jsc.fz-juelich.de/sdlrs/qmsvm}, 2023, last access on 15 Dec 2023.

\bibitem{cs_svm_code}
T.~Joachims, ``{SVM-Multiclass: Multi-Class Support Vector Machine},'' \url{https://www.cs.cornell.edu/people/tj/svm_light/svm_multiclass.html}, 2008, last access on 15 Dec 2023.

\bibitem{libsvm}
C.-C. Chang and C.-J. Lin, ``{LIBSVM: A library for support vector machines},'' \emph{ACM Transactions on Intelligent Systems and Technology}, vol.~2, pp. 27:1--27:27, 2011, software available at \url{http://www.csie.ntu.edu.tw/~cjlin/libsvm}.

\bibitem{toulouse_rs_dataset}
\BIBentryALTinterwordspacing
R.~Roscher, M.~Volpi, C.~Mallet, L.~Drees, and J.~D. Wegner, ``{SEMCITY TOULOUSE: A BENCHMARK FOR BUILDING INSTANCE SEGMENTATION IN SATELLITE IMAGES},'' \emph{ISPRS Annals of the Photogrammetry, Remote Sensing and Spatial Information Sciences}, vol. V-5-2020, pp. 109--116, 2020. [Online]. Available: \url{https://isprs-annals.copernicus.org/articles/V-5-2020/109/2020/}
\BIBentrySTDinterwordspacing

\bibitem{potsdam_rs_dataset}
ISPRS, ``{2D Semantic Labeling Contest - Potsdam},'' \url{https://www.isprs.org/education/benchmarks/UrbanSemLab/2d-sem-label-potsdam.aspx}, last access on 19 Dec 2023.

\bibitem{balanced_accuracy}
Scikit-learn, ``{Balanced accuracy},'' \url{https://scikit-learn.org/stable/modules/generated/sklearn.metrics.balanced_accuracy_score.html}, last access on 21 Dec 2023.

\bibitem{f1_score}
------, ``{F1 score},'' \url{https://scikit-learn.org/stable/modules/generated/sklearn.metrics.f1_score.html}, last access on 21 Dec 2023.

\end{thebibliography}

\begin{IEEEbiography}[{\includegraphics[width=1in,height=1.25in,clip,keepaspectratio]{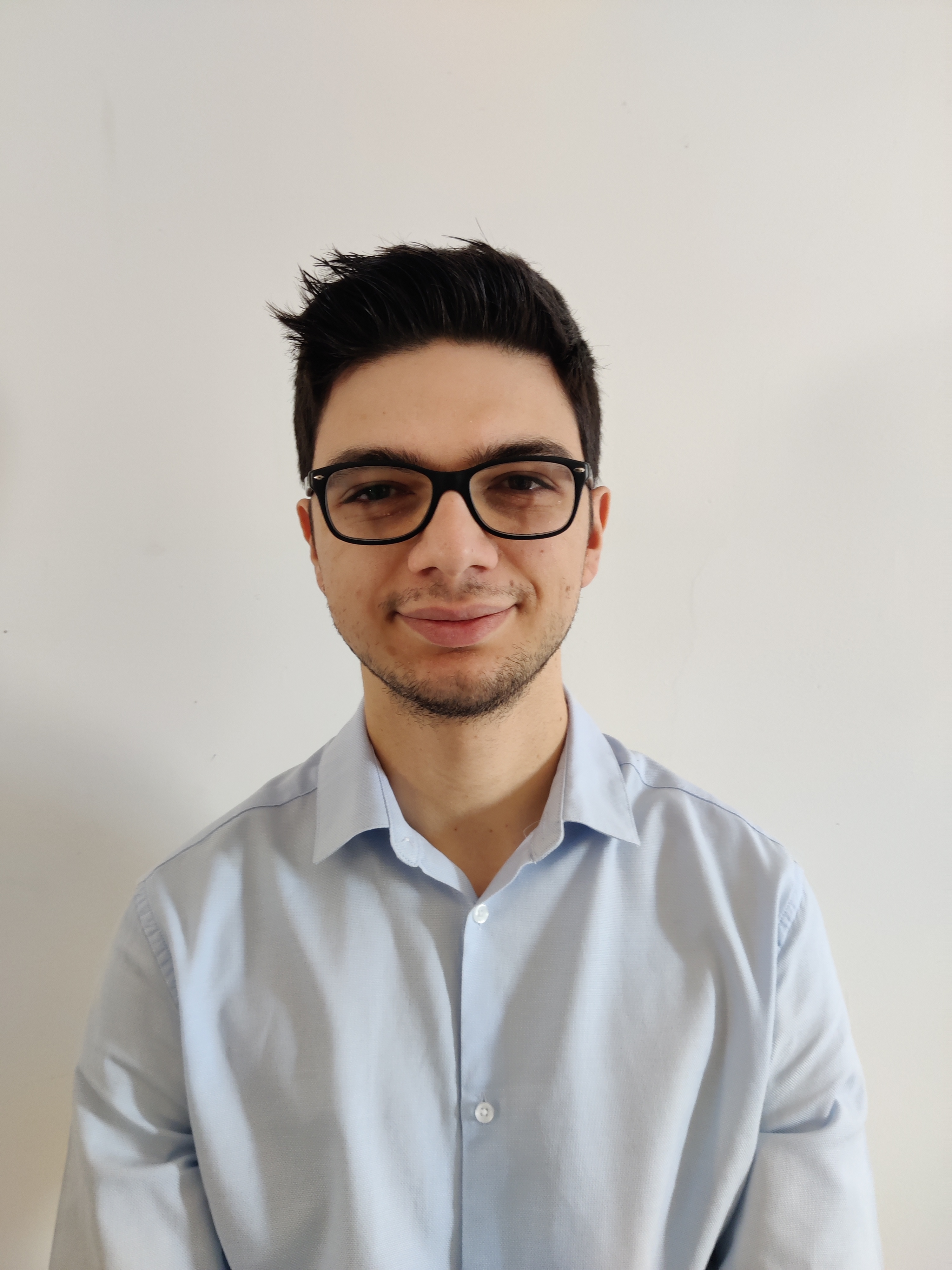}}]{Enrico Zardini}
received the B.Sc. and M.Sc. degrees in Computer Science from the University of Trento (Italy) in 2018 and 2020, respectively. He is currently pursuing the Ph.D. degree in Information and Communication Technology (ICT) at the University of Trento, where he is also enrolled in the Transdisciplinary Program in Quantum Science and Technology (QST). His research interests are quantum computing, quantum machine learning, and evolutionary algorithms.
\end{IEEEbiography}

\begin{IEEEbiography}
[{\includegraphics[width=1in,height=1.25in,clip,keepaspectratio]{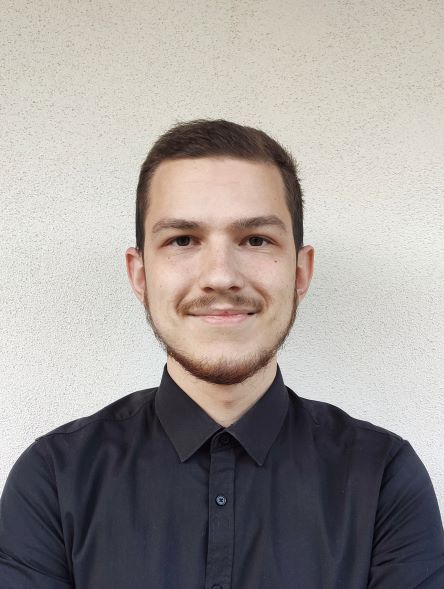}}]
{Amer Delilbasic} (Student Member, IEEE) received the B.Sc. and M.Sc. degrees in information and communication engineering from the University of Trento in 2019 and 2021, respectively. He is member of the ``AI and ML for Remote Sensing'' Simulation and Data Lab at the J\"{u}lich Supercomputing Centre, Forschungszentrum J\"{u}lich, Germany. He is currently pursuing the Ph.D. degree in computational engineering at the University of Iceland. He is an external researcher at $\Phi$-lab, European Space Agency, Frascati, Italy. His research interest is mainly in machine learning methods for remote sensing applications, with a particular focus on Quantum Computing (QC) and High Performance Computing (HPC).
\end{IEEEbiography}

\begin{IEEEbiography}[{\includegraphics[width=1in,height=1.25in,clip,keepaspectratio]{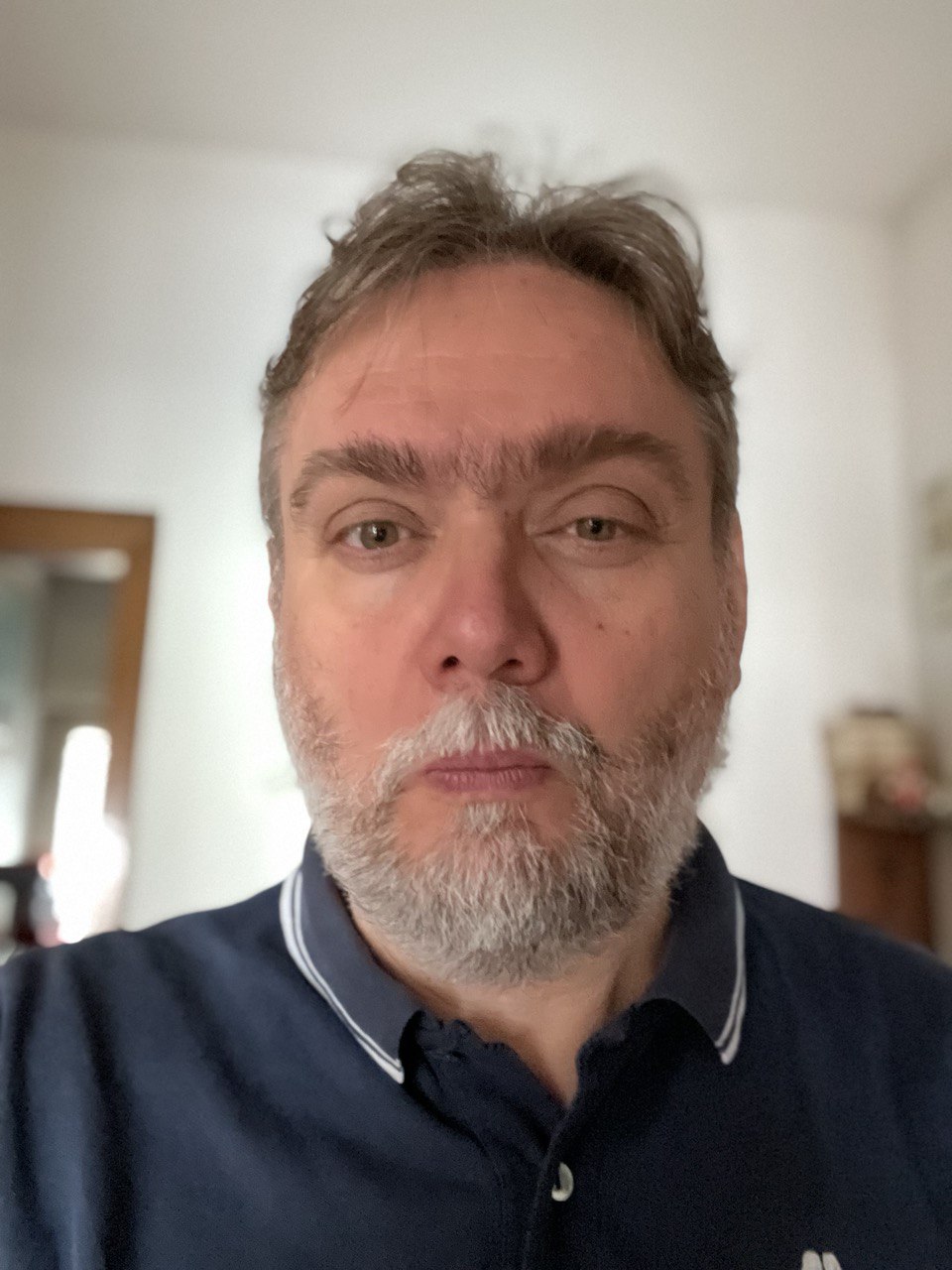}}]{Enrico Blanzieri} received the laurea degree (cum laude) in electronic engineering from the University of Bologna, Italy, and the PhD in cognitive science from the University of Turin, Italy, in 1992 and 1998, respectively. Since 2012, he is associate professor with the Department of Information Engineering and Computer Science (DISI), University of Trento, Italy where he works on machine learning, quantum computing and bioinformatics.
\end{IEEEbiography}

\begin{IEEEbiography}
[{\includegraphics[width=1in,height=1.25in,clip,keepaspectratio]{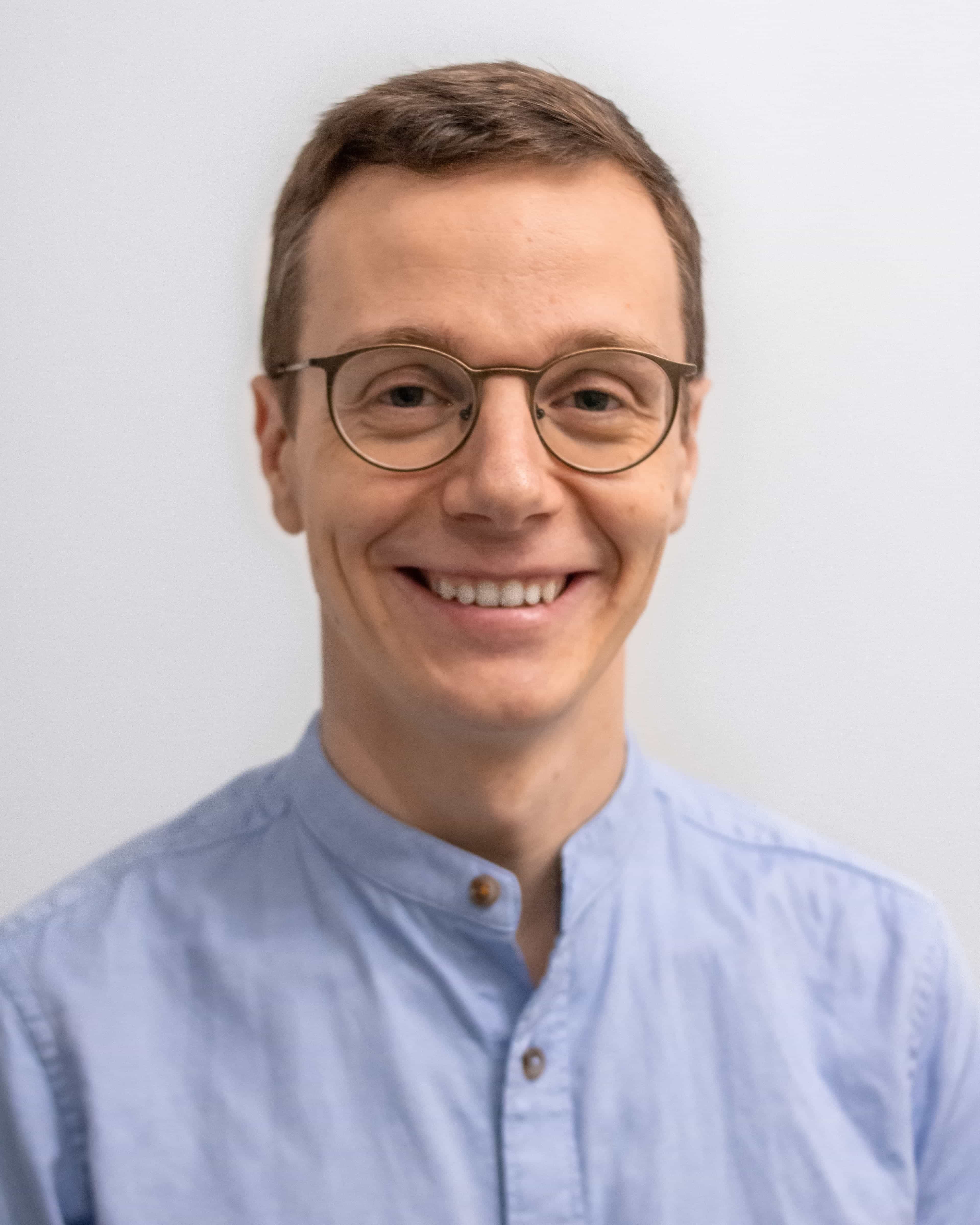}}]{Gabriele Cavallaro} (Senior Member, IEEE) received his B.Sc. and M.Sc. degrees in Telecommunications Engineering from the University of Trento, Italy, in 2011 and 2013, respectively, and a Ph.D. degree in Electrical and Computer Engineering from the University of Iceland, Iceland, in 2016. From 2016 to 2021 he has been the deputy head of the ``High Productivity Data Processing'' (HPDP) research group at the J\"{u}lich Supercomputing Centre (JSC), Forschungszentrum J\"{u}lich, Germany. Since 2022, he is the Head of the ``AI and ML for Remote Sensing'' Simulation and Data Lab at the JSC and an Adjunct Associate Professor with the School of Natural Sciences and Engineering, University of Iceland, Iceland. From 2020 to 2023, he held the position of Chair for the High-Performance and Disruptive Computing in Remote Sensing (HDCRS) Working Group under the IEEE GRSS Earth Science Informatics Technical Committee (ESI TC). In 2023, he took on the role of Co-chair for the ESI TC. Concurrently, he serves as Visiting Professor at the $\Phi$-lab within the European Space Agency (ESA), where he contributes to the Quantum Computing for Earth Observation (QC4EO) initiative. Additionally, he has been serving as an Associate Editor for the IEEE Transactions on Image Processing (TIP) since October 2022. He was the recipient of the IEEE GRSS Third Prize in the Student Paper Competition of the IEEE International Geoscience and Remote Sensing Symposium (IGARSS) 2015 (Milan - Italy). His research interests include remote sensing data processing with parallel machine learning algorithms that scale on distributed computing systems and innovative computing technologies.
\end{IEEEbiography}

\begin{IEEEbiography}[{\includegraphics[width=1in,height=1.25in,clip,keepaspectratio]{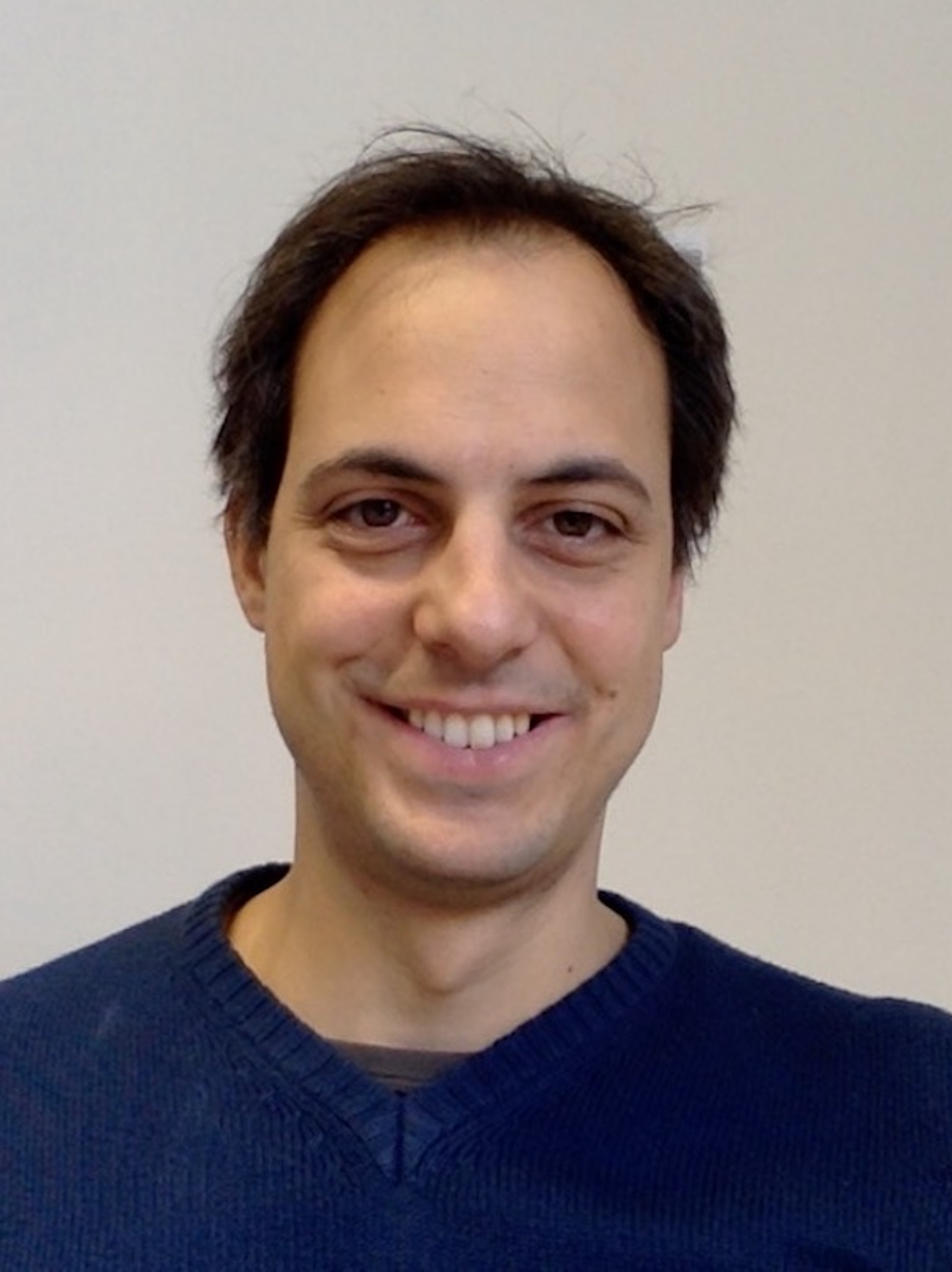}}]{Davide Pastorello} received the M.Sc. degrees in Physics (cum laude) and the Ph.D. degree in Mathematics from the University of Trento in 2011 and 2014, respectively. From 2015 to 2019 he was postdoc researcher at Deptartment of Mathematics, University of Trento, also with a 2-year grant from Fondazione Caritro as P.I. of a project on quantum computing. From 2020 to 2023 he was Assistant Professor at the Department of Information Engineering and Computer Science (DISI), University of Trento. Since 2023, he is Assistant Professor at Deptartment of Mathematics, University of Bologna.  His main research interests are: mathematical foundations of quantum mechanics, mathematical methods in quantum computing and quantum machine learning.
\end{IEEEbiography}

\vfill

\end{document}